# Inertia-Constrained Generation Scheduling: Sample Selection, Learning-Embedded Optimization Modeling, and Computational Enhancement

Mingjian Tuo, *Member*, *IEEE*, Fan Jiang, *Student Member*, *IEEE*, Xingpeng Li, *Senior Member*, *IEEE*, and Pascal Van Hentenryck, *Senior Member*, *IEEE*

***Abstract*— Day-ahead generation scheduling is conducted by solving security-constrained unit commitment (SCUC) problem. Fast-growing inverter-based resources dramatically reduces grid inertia, compromising system dynamic stability. Traditional SCUC (T-SCUC), without any inertia requirements, may no longer be effective for renewables-dominated grids. To address this, we propose the active linearized sparse neural network-embedded SCUC (ALSNN-SCUC) model, utilizing machine learning (ML) to incorporate system dynamic performance. A multi-output deep neural network (DNN) model is trained offline on strategically-selected data samples to accurately predict frequency stability metrics: locational RoCoF and frequency nadir. Structured sparsity and active ReLU linearization are implemented to prune redundant DNN neurons, significantly reducing its size while ensuring prediction accuracy even at high sparsity levels. By embedding this ML-based frequency stability predictor into SCUC as constraints, the proposed ALSNN-SCUC model minimizes its computational complexity while ensuring frequency stability following *G*-1 contingency. Case studies show that the proposed ALSNN-SCUC can enforce pre-specified frequency requirements without being overly conservative, outperforming five benchmark models including T-SCUC, two physics-based SCUC, and two ML-based SCUC. The proposed sparsification and active linearization strategies can reduce the DNN-SCUC computing time by over 95% for both IEEE 24-bus and 118-bus systems, demonstrating the effectiveness and scalability of the proposed ALSNN-SCUC model.***

## NOMENCLATURE

*Sets*

| | |
|---|---|
| $G$ | Set of generators. |
| $G(n)$ | Set of generators on bus $n$. |
| $K$ | Set of lines. |
| $K^{+}(n)$ | Set of lines with bus $n$ as receiving bus. |
| $K^{-}(n)$ | Set of lines with bus $n$ as sending bus. |
| $T$ | Set of time periods. |
| $N$ | Set of buses. |
| $NS$ | Set of samples. |
| $NL$ | Set of neural network layers. |
| $\mathcal{H}$ | Set of active selected neurons. |
| $\bar{\mathcal{H}}$ | Set of not selected neurons. |

*Indices*

| | |
|---|---|
| $g$ | Generator $g$. |
| $k$ | Line $k$. |
| $t$ | Time $t$. |
| $n$ | Bus $n$. |
| $q$ | Neural network layer $q$. |
| $l$ | $l$-th neuron of a neural network layer. |
| $s$ | Sample $s$. |

*Parameters*

| | |
|---|---|
| $c_g$ | Linear operation cost for generator $g$. |
| $P_g^{min}$ | Minimum output limit of generator $g$. |
| $P_g^{max}$ | Maximum output limit of generator $g$. |
| $P_k^{max}$ | Long-term thermal line limit for line $k$. |
| $b_k$ | Susceptance of line $k$. |
| $D_{n,t}$ | Forecasted demand at bus $n$ in period $t$. |
| $E_{n,t}$ | Forecasted renewable generation at bus $n$ in period $t$. |
| $R_g^{hr}$ | Ramping limit of generator $g$. |
| $R_g^{re}$ | Reserve ramping limit of generator $g$. |
| $c_g^{NL}$ | No load cost for generator $g$. |
| $c_g^{SU}$ | Startup cost of generator $g$. |
| $c_g^{RE}$ | Reserve cost of generator $g$. |
| $DT_g$ | Minimum down time of generator $g$. |
| $UT_g$ | Minimum on time of generator $g$. |
| $nT$ | Number of time periods. |
| $A$ | A big real number. |
| $RoCoF_{\mathrm{lim}}$ | Pre-specified RoCoF threshold. |
| $f_{\mathrm{nom}}$ | System frequency nominal value. |
| $f_{\mathrm{lim}}$ | Pre-specified minimal frequency threshold. |
| $W_q$ | Weights matrix of layer $q$. |
| $b_q$ | Bias matrix of layer $q$. |
| $W_{N_L+1}^{dev}$ | Weights matrix of last layer for maximal frequency deviation prediction. |
| $W_{N_L+1}^{rcf}$ | Weights matrix of last layer for maximal RoCoF prediction. |
| $b_{N_L+1}^{dev}$ | Bias matrix of last layer for maximal frequency deviation prediction. |
| $b_{N_L+1}^{rcf}$ | Bias matrix of last layer for maximal RoCoF prediction. |
| $\mathrm{UB}_{q(l)}$ | Upper bound of preactivated value of $l$-th neuron of layer $q$. |
| $\mathrm{LB}_{q(l)}$ | Lower bound of preactivated value of $l$-th neuron of layer $q$. |

*Variables*

| | |
|---|---|
| $P_{g,t}$ | Output of generator $g$ in period $t$. |
| $r_{g,t}$ | Reserve from generator $g$ in period $t$. |
| $u_{g,t}$ | Commitment status of generator $g$ in period $t$. |

Mingjian Tuo is with Hubei Key Laboratory of Energy Storage and Power Battery (Hubei University of Automotive Technology), Shiyan, Hubei, 442000, CN (e-mail: mtuo@huat.edu.cn). He was with the University of Houston. Fan Jiang and Xingpeng Li are with the Department of Electrical and Computer Engineering, University of Houston, Houston, TX, 77204, USA (e-mail: fjiang6@uh.edu; xingpeng.li@asu.edu). Pascal Van Hentenryck is with the H. Milton Stewart School of Industrial and Systems Engineering, Georgia Institute of Technology, Atlanta, GA, 30332, USA (email: pascal.vanhentenryck@isye.gatech.edu).

| | |
|---|---|
| $v_{g,t}$ | Start-up variable of generator $g$ in period $t$. |
| $P_{k,t}$ | Flow online $k$ in period $t$. |
| $\theta_{n,t}$ | Phase angle of bus $n$ in period $t$. |
| $\theta_{m,t}$ | Phase angle of bus $m$ in period $t$. |
| $\mu_{g,t}$ | Maximal output indicator of generator $g$ in period $t$. |
| $\rho_{g,t}$ | Maximal output value of generator $g$ in period $t$. |
| $z_{q(l),t}$ | Preactivated value of $l$-th neuron of layer $q$ in period $t$. |
| $\hat{z}_{q(l),t}$ | Activated value of $l$-th neuron of layer $q$ in period $t$. |
| $a_{q(l),t}$ | Activation status $l$-th neuron's ReLU of layer $q$ in period $t$. |

# I. Introduction

As modern power systems increasingly incorporate renewable energy sources (RES), the system configurations and generation mix have changed substantially, leading to reduced system inertia, altered dynamic behaviors, and reduced grid operation regions. Considerable emerging challenges are to enforce power system dynamic performance such as system frequency stability, a vital aspect for the dependable functioning of the grids [1]-[2]. Conventionally, power systems have relied on the inertia provided by large-scale synchronous generators, which plays a vital role in stabilizing the grid against frequency fluctuations [3]. However, as RESs such as wind and solar power, which are typically connected through power electronic converters, increasingly displace synchronous generators, the inherent grid inertia diminishes [4]. This transition has led to notable incidents such as the blackout in South Australia in 2016 [5] and the British blackout in 2019 [6], both of which were linked to issues with frequency control.

Security-constrained unit commitment (SCUC) is solved for power system day-ahead generation scheduling. Traditional SCUC optimizes generator status only to ensure steady-state grid reliability by procuring sufficient generating resources, which is unable to ensure system dynamics and thus not effective for future RES-dominated power grids. Addressing the aforementioned challenges has led to the strategic inclusion of frequency security requirements within the SCUC framework. This would enable grid operators to assess the instability risk and automatically adjust the dispatch of resources to mitigate the impact of frequency deviations [7]. For instance, EirGrid in Ireland has set a specific constraint for synchronous inertial response, ensuring that the inertia level does not fall below a certain threshold of 23 GWs [8]. Similarly, the transmission system operator in Sweden directed a nuclear facility to reduce its generation by 100 MW as a precaution to minimize the risk of disturbances emanating from the plant [9].

Prior efforts in the literature that consider frequency dynamics are primarily focused on average system frequency (ASF)-based and center of inertia (COI)-based models. The SCUC model in [10] includes model-based constraints to secure system frequency stability. References [11]-[12] enhance the conventional frequency response model by integrating converter control strategies and incorporating these into SCUC with extra constraints limiting the rate of change of frequency (RoCoF). Yet, by averaging the inertia and assuming uniform frequency response across the entire system, COI-based models may not accurately capture local variations in frequency behavior, particularly for large geographically-dispersed power systems. This can lead to inadequate responses to local disturbances, potentially resulting in localized instability issues or the need for excessive spinning reserves. Research in [13] explores the dynamics across multiple regions, proposing a model that captures the evolution of the center of inertia with specific inter-area oscillations. Another study of [14] investigates the geographical differences in the spatial distribution of frequency. However, such physical model-based frequency constraints often assume linear or quasi-linear relationships within system dynamics, possibly leading to inadequate or overly conservative operational decisions, limiting the grid efficiency and flexibility or reducing grid reliability.

A novel development in the field, highlighted in [15], adopts a data-driven method. This method embeds frequency stability-related constraints derived from deep neural networks (DNN) into SCUC. Additionally, [16] introduces a data-driven approach that incorporates forward propagation equations into SCUC by translating them into a set of mixed-integer linear constraints for frequency security, which also defines extra binary variables. Although these efforts present promising directions, they come with the following challenge: integrating DNN-based constraints into SCUC can dramatically increase the computational demands, potentially making the optimization task unsolvable [17]. The deterioration in computational efficiency stemming from integrating DNNs into the SCUC resolution process necessitates further investigation.

To bridge the aforementioned gaps, this paper proposes a machine learning (ML)-assisted active linearized sparse neural network-based SCUC (ALSNN-SCUC) model. The main contributions of this work are summarized as follows:

- The proposed ALSNN-SCUC is developed by adopting a learning-embedded optimization modeling strategy. It enhances a traditional SCUC by incorporating additional frequency stability constraints, derived from a ML-based locational RoCoF and frequency nadir predictor. This ML model can predict multiple frequency stability metrics jointly with minimum size and has the potential to be extended to capture other stability performance metrics.
- A model-based sample generation method is developed to efficiently create reliable training datasets. Compared to arbitrary data generation methods, it can significantly reduce the divergence rates of generated grid conditions and enable an efficient sample creation process. The concept of off-line training can also handle datasets with large amounts of training cases, increasing the accuracy and robustness of trained frequency tracking model.
- A novel sample selection strategy is also designed to select representative samples to ensure the trained ML model is generally accurate and also robust for scenarios when the frequency nadir and RoCoF are very close to prespecified tolerances. This strategy improves the ML model and thus the ALSNN-SCUC model to better capture the frequency stability performance.
- Multiple computational enhancement strategies are developed to ensure the proposed ALSNN-SCUC can be efficiently solved, while existing work simply embeds full-complexity dense DNN into SCUC, leading to pro-hibitive computational burdens due to dense parameter matrices. Sparse technology is adopted to reduce the size and complexity of the ML model until desire prediction accuracy cannot be achieved. Then, it will be actively linearized on strategically-selected neurons; particularly, the activation function of rectified linear unit (ReLU) will be linearized. This strategy will substantially reduce the constraints and variables including binary variables required for ALSNN-SCUC that embeds the ML model. Compared to the DNN-SCUC model, The ALSNN-SCUC model achieves a 99% reduction in solving time on the IEEE 24-bus system and a 95% reduction in solving time on the IEEE 118-bus system

The rest of this paper is organized as follows: Section II describes both model-based and data-driven solution methods. Section III establish the ML-based frequency constraints. Section IV explains the active ReLU linearization of sparse neural network. Section V presents the proposed ALSNN-SCUC model. Results and analysis are presented in Section VI, while Section VII concludes with final remarks and future work.

## II. Overview of Solution

In this section, we first present the process for establishing constraints based on model requirements for frequency (such as restrictions on the RoCoF and minimum frequency levels), followed by an explanation on the data-driven formulation and the creation of datasets through a model-oriented approach.

### *A. Frequency Dynamics Model*

Power system frequency is a key indicator of the grid dynamic performance. Related control was usually developed through simplified models, either by considering the entire system as a singular bus or through the COI concept [18]-[19]. The relationship between power and frequency can be described by the following swing equation in (1),

$$\Delta P_m - \Delta P_e = M\frac{d\Delta\omega}{dt} + D\Delta\omega \tag{1}$$

where $\Delta P_m$ is the mechanical power deviation; and $\Delta P_e$ is the change in electric power; $M = 2H$ represents the normalized inertia constant; $D$ denotes the damping constant. Considering a power disruption $\Delta P$, the initial RoCoF requirement for the uniform system model is determined by (2),

$$f_{rcf}(\Delta P) = \frac{\Delta P}{2HS_B}\omega_n \leq -\mathrm{RoCoF_{lim}} \tag{2}$$

However, only focusing on the COI for frequency changes may fail to address local frequency variations that could destabilize the power system [13]. Thus, a model that applies swing equations to every individual bus is introduced to precisely capture the oscillatory patterns across the network [20],

$$m_i\ddot{\theta}_i + d_i\dot{\theta}_i \;=\; P_{in,i} - \sum_{j=1}^{n} b_{ij}\,sin\,(\theta_i \;-\; \theta_j) \tag{3}$$

where $P_{in,i}$ denoting the power input. The dynamic equation describing the phase angle $\theta$ of generator buses is formulated by emphasizing the influence of network connectivity on the nodal dynamics of the power system [21],

$$M\ddot{\theta} \;+\; D\,\dot{\theta} \;=\; P - L\,\theta \tag{4}$$

where $L$ is the Laplacian matrix. The RoCoF value at bus $i$ can then be derived [14], [21],

$$f_{rcf,i}(\Delta P, t) = \frac{\Delta P e^{-\frac{\gamma t}{2}}}{2\pi m}\sum_{\alpha=1}^{N_g}\frac{\beta_{\alpha i}\beta_{\alpha b}}{\sqrt{\frac{\lambda_\alpha}{m}-\frac{\gamma^2}{4}}\Delta t}\begin{bmatrix} e^{-\frac{\gamma\Delta t}{2}}\,sin\left(\sqrt{\frac{\lambda_\alpha}{m}-\frac{\gamma^2}{4}}(t+\Delta t)\right) \\ -\,sin\left(\sqrt{\frac{\lambda_\alpha}{m}-\frac{\gamma^2}{4}}\,t\right)\end{bmatrix} \tag{5}$$

### *B. Framework of Data-driven Approach*

The foundational framework of the SCUC model is outlined in equations (6) and (7), establishing the operational condition of the power systems within the scheduling period. Equation (7) seeks to reduce total operational expenses, subject to a range of constraints related to system operations.

$$min.\quad \mathcal{C}(s_t, u_t) \tag{6}$$

$$s.t.\quad \mathcal{F}(s_t, u_t, d_t, r_t) = 0, \mathcal{G}(s_t, u_t, d_t, r_t) \leq 0, \forall t \tag{7}$$

where $\mathcal{F}$ and $\mathcal{G}$ are the equality and inequality constraints respectively, including power balance, generators related limits, network related constraints and frequency related constraints; $s_t$ denotes the power system states, and $u_t$ is the generation output power at period $t$. $d_t$ and $r_t$ denote the predicted load and renewable generation profiles respectively.

Integrating constraints pertinent to frequency security, which are informed by the frequency response model, into the SCUC, plays an important role in ensuring the stability of system frequency following a potential power perturbation $\varpi_t$. The essence of transitioning to a data-driven ML modeling method is to replace physical model-based constraints with DNN formulations, aiming to mitigate approximation errors that may lead to overly conservative or insecure outcomes,

$$\hat{h}^f(s_t, u_t, r_t, \varpi_t) \leq \varepsilon \tag{8}$$

where the vector $\varepsilon$ represents predefined threshold requirements. The nonlinear DNN-based predictor provides two outputs in this paper: maximal frequency deviation and maximal locational RoCoF. Although this work focuses on frequency stability, it is worth noting that this ML-based method can be generalized to provide additional outputs for dynamic performance metrics in other stability categories such as maximum rotor angle difference and largest eigenvalues.

Generators, especially large thermal units, contribute significantly to the inertia of the power grid, which is crucial for resisting changes in system frequency. When a generator goes offline unexpectedly, the system's inertia decreases sharply, reducing its ability to buffer against rapid frequency fluctuations. This sudden reduction in inertia can lead to rapid and significant frequency drops, posing a substantial risk to the stability and reliability of the power network. Consequently, the loss of a generator not only challenges the balance of supply and demand but also undermines the system's structural ability to maintain stable frequency. This study identifies the scenario of a single generator failure, known as the G-1 event, as the most critical contingency examined.

The overview of the working pipeline is shown in Fig. 1:

1. DNN-based predictor is trained with sparse computation on a dataset generated by three SCUC variants. This dataset captures diverse operational scenarios, in-cluding $G$-1 contingencies, to ensure the DNN learns critical inertia-related dynamics.

2. Inertia constraints are enforced via predicted frequency stability metrics (RoCoF and frequency nadir), which are embedded as additional constraints in SCUC.
3. The nonlinear ReLU activations in the DNN are linearized on a subset of neurons using an active selection strategy.

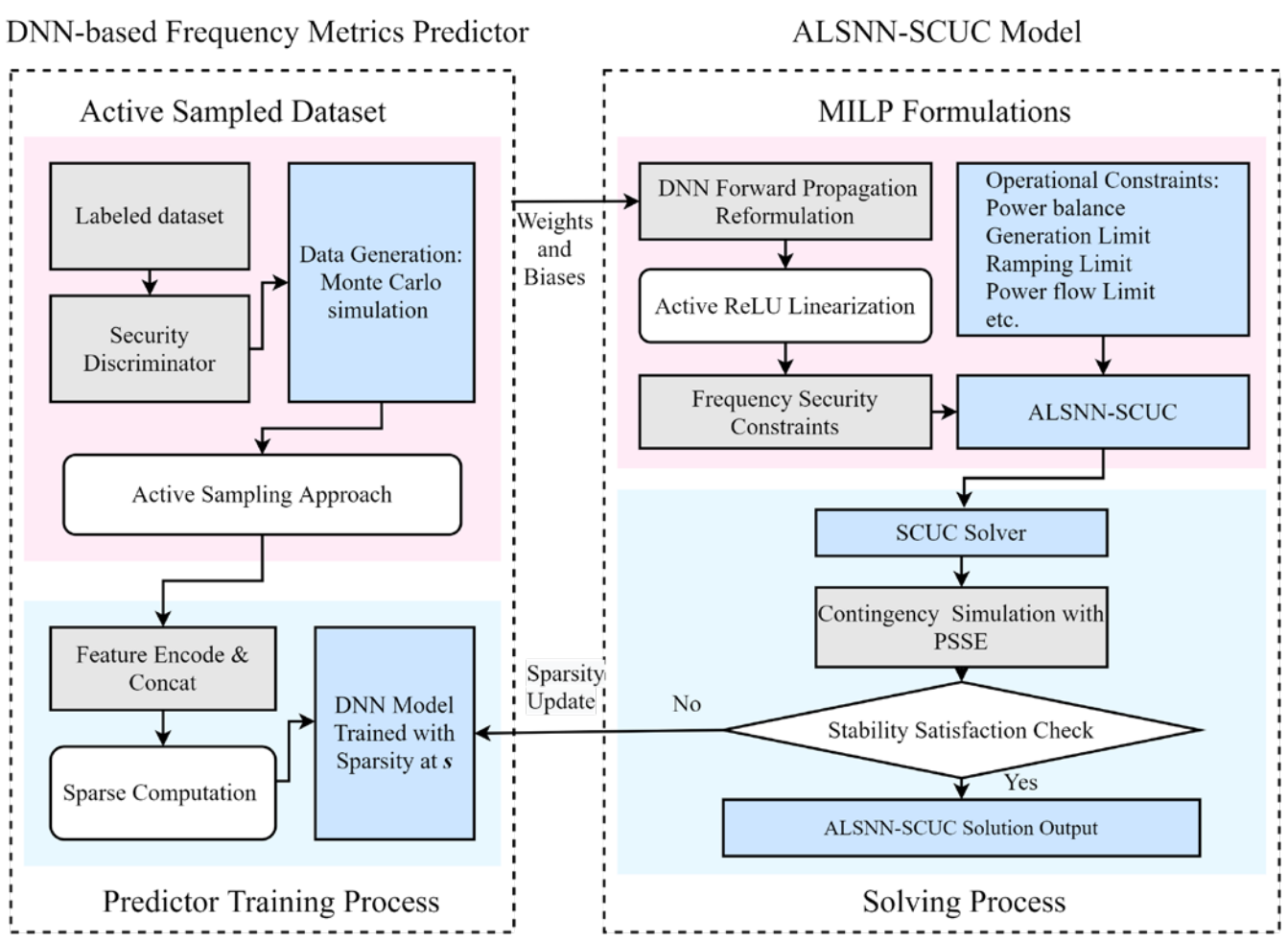

Fig. 1. Overview of the proposed approach.

### C. *Model-based Data Generation*

The use of randomized power injections [15] may enhance reliability across varied operational settings for specific generators. Nonetheless, in practical power systems with many generators, the complexity added by random injection methods expands the state space considerably. This expansion can trigger instability early in simulation processes, potentially undermining the stability of the entire system when faced with unexpected contingencies [22]. Thus, the indiscriminate application of random injections might lead to early divergences in simulations, posing significant risks to system stability during unforeseen events.

Different from the less predictable method of generating random data, our strategy leverages a model-based approach that produces training datasets. The proposed methodology utilizes three specialized SCUC models: (i) traditional SCUC without any frequency stability constraints (T-SCUC), (ii) system equivalent RoCoF-constrained SCUC (ERC-SCUC), and (iii) locational RoCoF-constrained SCUC (LRC-SCUC). T-SCUC generates baseline data for normal operations, while ERC-SCUC and LRC-SCUC impose progressively stricter frequency security requirements. These models are used to systematically generate data through Monte Carlo simulations, covering a wide array of load conditions and RES scenarios. The central objective is to minimize the overall operational costs, which includes managing fuel usage, idle capacities, startup processes, and maintaining adequate reserves. Specifically, the ERC-SCUC model incorporates system-wide uniform RoCoF constraint shown in (9),

$$f_{rcf}(\varpi_t) \leq -\mathrm{RoCoF}_{\mathrm{lim}} \tag{9}$$

Equations (10)-(11) are derived from (4)-(5), presenting locational RoCoF constraints.

$$f_{rcf,i}(\varpi_t, t_1) \leq -\mathrm{RoCoF}_{\mathrm{lim}}, \forall\, i \in N_G, \forall\, t \tag{10}$$

$$f_{rcf,i}(\varpi_t, t_2) \leq -\mathrm{RoCoF}_{\mathrm{lim}}, \forall\, i \in N_G, \forall\, t \tag{11}$$

Table I listed three separate benchmark SCUC models, each characterized by unique objective functions and a set of specific constraints they adhere to. It's important to note that the constraints governing frequency dynamics exhibit a nonlinear nature. In order to integrate these nonlinear constraints within the framework of Mixed-Integer Linear Programming (MILP), a strategy involving piecewise linear approximation is utilized. For a detailed explanation of all models, please refer to [14].

TABLE I. Benchmark SCUC models for data sample generation and comparison with the proposed ML-assisted ALSNN-SCUC model

| Model | Objective Function | Shared Sets of Constraints | Unique Constraints |
|---|---|---|---|
| T-SCUC | | | N/A |
| ERC-SCUC | (6) | (7) | (9) |
| LRC-SCUC | | | (10)-(11) |

## III. DEEP LEARNING-BASED FREQUENCY CONSTRAINTS

In this section, we first describe the definition of input feature vector for the DNN-based predictor. Then the architecture of DNN is designed for multi-states tracking. Finally, the forward propagation equations of the DNN are reformulated into MILP format for SCUC model incorporation.

### A. *Power System Feature Definition*

The integration of RES introduces significant variability and complexity, particularly influencing system frequency dynamics. The effectiveness of models designed to predict RoCoF and the lowest frequency nadir depends heavily on a thorough incorporation of specific power system characteristics, including the magnitude and location of disturbances, the operational status of generators, the strategies for unit dispatch, and overall grid conditions. Each of these factors plays a crucial role in the system's response to changes and must be accurately represented in the training datasets used for developing predictive models [23]. For a case at hour $t$, the generator status vector is defined as follow,

$$u_t = [u_{1,t}, u_{2,t}, \cdots, u_{NG,t}] \tag{12}$$

The configuration of the disturbance feature vector is designed to capture the effects of the most significant generation outage. Such an approach enables an accurate depiction of the system contingency magnitude.

$$P_t^{con} = \max_{g \in G}(P_{1,t}, P_{2,t}, \cdots, P_{NG,t}) \tag{13}$$

The disturbance location is identified by the index corresponding to the generator contributing the highest power output. Information regarding both the magnitude and the specific location of the disturbance is then integrated into the disturbance feature vector, following methodologies proposed in earlier study [15],

$$g_t^{con} = arg \max_{g \in G}(P_{1,t}, P_{2,t}, \cdots, P_{NG,t}) \tag{14}$$

$$\varpi_t^G = [0, \cdots, 0, \underbrace{P_t^{con}}_{g_t^{con} th\ element}, 0, \cdots, 0] \tag{15}$$

The Laplacian matrix $L$ of the grid and the Fiedler mode value show a degree of correlation with power-angle properties, which are affected by injections of active power [14]. The active power contributions from each synchronous generator are compiled into a composite feature vector for analysis.

$$P_t = [P_{1,t}, \cdots, P_{2,t}, \cdots, P_{NG,t}] \tag{16}$$

And the overall feature vector of a case $x_t$ at hour $t$ can be defined as,

$$x_t = [u_{1,t}, \cdots, u_{NG,t}, \ \varepsilon_{1,t}, \cdots, \ \varepsilon_{NG,t}, P_{1,t}, \cdots, P_{NG,t}] \tag{17}$$

Training DNNs with feature vector $x_t$ that encapsulate all critical factors provides substantial advantages for frequency stability prediction in power system.

### *B. Formulation of DNN-based Frequency Metrics Predictor*

Leveraging a rich dataset enables the training of a DNN to monitor system frequency dynamics and other critical metrics accurately. The proposed predictor can capture complex relationships and patterns that may be challenging to represent accurately with traditional mathematical models. As a result, utilizing constraints formulated from the DNN forward propagation provides a more nuanced and robust understanding of the frequency dynamics within power systems, particularly valuable in scenarios marked by complex, high-dimensional, or nonlinear dynamics. This advantage is crucial for the precise tracking and modeling of the dynamic states within power systems, highlighting the DNN's potential to enhance operational reliability and efficiency significantly.

The architecture of frequency metrics predictor $\hat{h}$ is shown as in Fig. 2, the function is constructed as,

$$\hat{h}(x_t, W, b) = \begin{bmatrix} \chi_1 \\ \chi_2 \\ \vdots \\ \chi_n \end{bmatrix} \tag{18}$$

where $W$ and $b$ represent the well-trained weights and biases for frequency metrics predictor, $\chi_n$ denotes the prediction of frequency dynamic metrics. The model consolidates functions through the introduction of an output layer with multiple outputs. This approach significantly reduces the size and number of needed DNN model, thereby reducing the total number of additional variables introduced into the SCUC formulations.

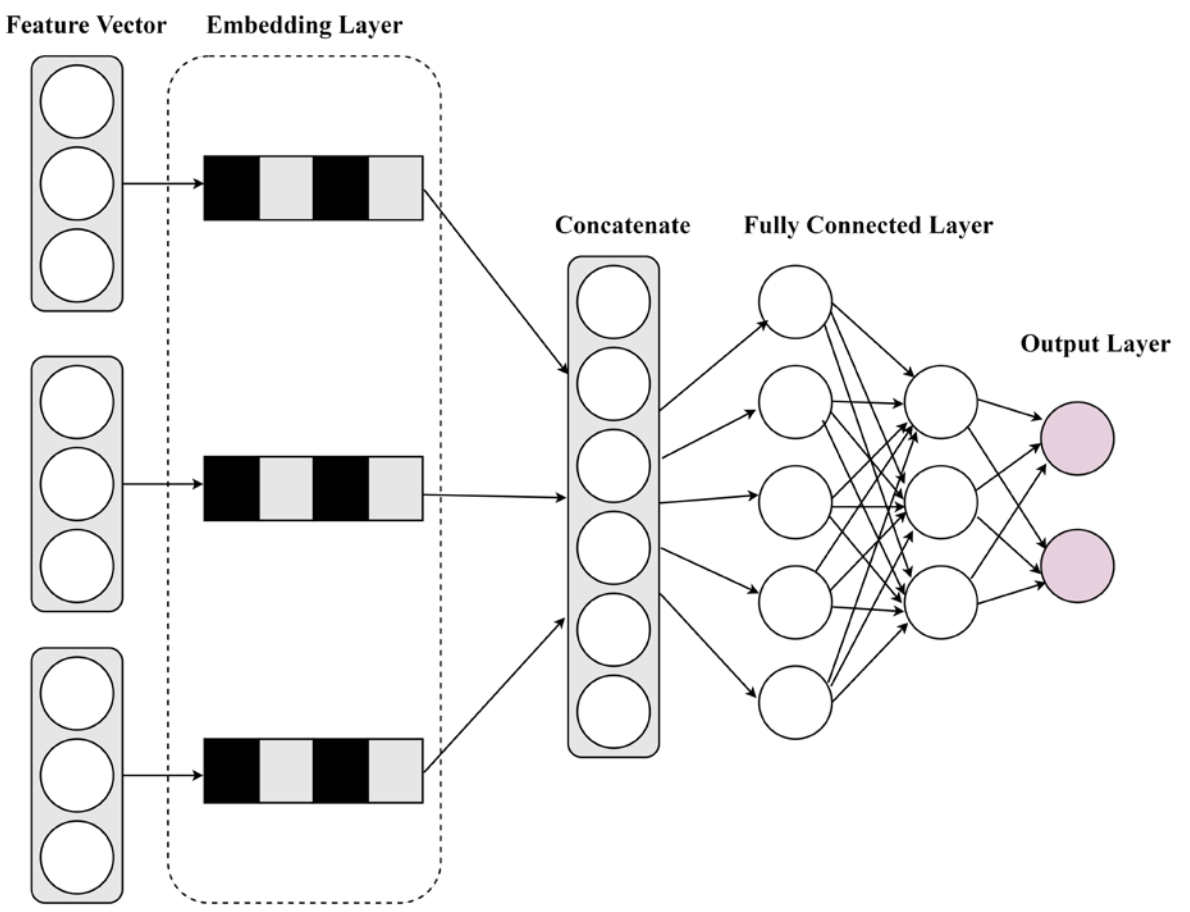

Fig. 2. Example of deep neural network-based predictor.

For a DNN with $N_L$ hidden layers employing ReLU as the activation function for each layer, the output layer is configured as a fully connected layer, yielding the output vector $\chi$. Within this output vector, $\hat{f}_{dev}$ and $\hat{f}_{rcf}$ represent the predicted maximum frequency deviation and maximal RoCoF values, respectively, for worst-case at hour $t$. The forward propagation equations of predictor $\hat{h}$ are as follows,

$$\hat{z}_1 = x_t W_1 + b_1 \tag{19}$$

$$\hat{z}_q = z_{q-1} W_q + b_q \tag{20}$$

$$z_q = \max(\hat{z}_q, 0) \tag{21}$$

$$\hat{f}_{dev} = z_{N_L} W_{N_L+1}^{dev} + b_{N_L+1}^{dev} \tag{22}$$

$$\hat{f}_{rcf} = z_{N_L} W_{N_L+1}^{rcf} + b_{N_L+1}^{rcf} \tag{23}$$

In the training phase, the objective function is to minimize the mean squared error across all samples. The error represents the average of the squares of the differences between each predicted output and its corresponding actual label, outlined as follows,

$$\min_{\Phi} \frac{1}{N_S} \sum_{s=1}^{N_S} \left(\Delta f_{max} - \hat{f}_{dev}\right)^2 + \left(\dot{f}_{max} - \hat{f}_{rcf}\right)^2 \tag{24}$$

where $\Phi = \{W_q, b_q, W_{N_L+1}^{dev}, W_{N_L+1}^{rcf}, b_{N_L+1}^{dev}, b_{N_L+1}^{rcf}\}$ represent the set of optimization variables.

Due to the inherent nonlinearity of ReLU function, conventional commercial optimization solvers face challenges in directly solving the SCUC model. To incorporate these nonlinear constraints into the mixed integer linear programming (MILP) model, binary variables $a_{q[l]}$ are introduced, which represent the activation status of the ReLU unit at $l$th neuron of $q$th layer.

$$z_{q[l],t} \leq \hat{z}_{q[l],t} + A(1 - a_{q[l],t}), \forall q, \forall l, \forall t, \tag{25a}$$

$$z_{q[l],t} \geq \hat{z}_{q[l],t}, \forall q, \forall l, \forall t, \tag{25b}$$

$$z_{q[l],t} \leq A a_{q[l],t}, \forall q, \forall l, \forall t, \tag{25c}$$

$$z_{q[l],t} \geq 0, \forall q, \forall l, \forall t, \tag{25d}$$

$$a_{q[l],t} \in \{0, 1\}, \forall q, \forall l, \forall t, \tag{25e}$$

## IV. Active Linearization of Sparse Neural Network

Completely reformulation of the DNN avoids introducing approximation errors, this linearization process does bring in a dense parameter matrix and additional binary variables into the MILP model. Consequently, the computational burden of the SCUC model would notably increase, particularly when multiple scheduling periods are constrained. In this section, we utilize sparse computation and active ReLU linearization to achieve optimal efficiency.

### *A. Sparse Computation*

Embedding a DNN-based model into MILP formulations significantly increases computational demands. Such surge in complexity primarily arises from the large number of parameters and the dense connections typical of DNNs, which intensify the computational load and memory requirements during optimization.

To manage these challenges, sparse computation techniques are applied which strategically reduces the density of network parameters by selectively pruning unnecessary connections and weights, thereby streamlining the computational process. The approach not only reduces the computational burden but also enhances the overall efficiency of solving MILP problems, maintaining high levels of accuracy in predictions while ensuring that the model remains computationally manageable [17]. The example of training algorithm for sparse computation is shown in Fig. 3.

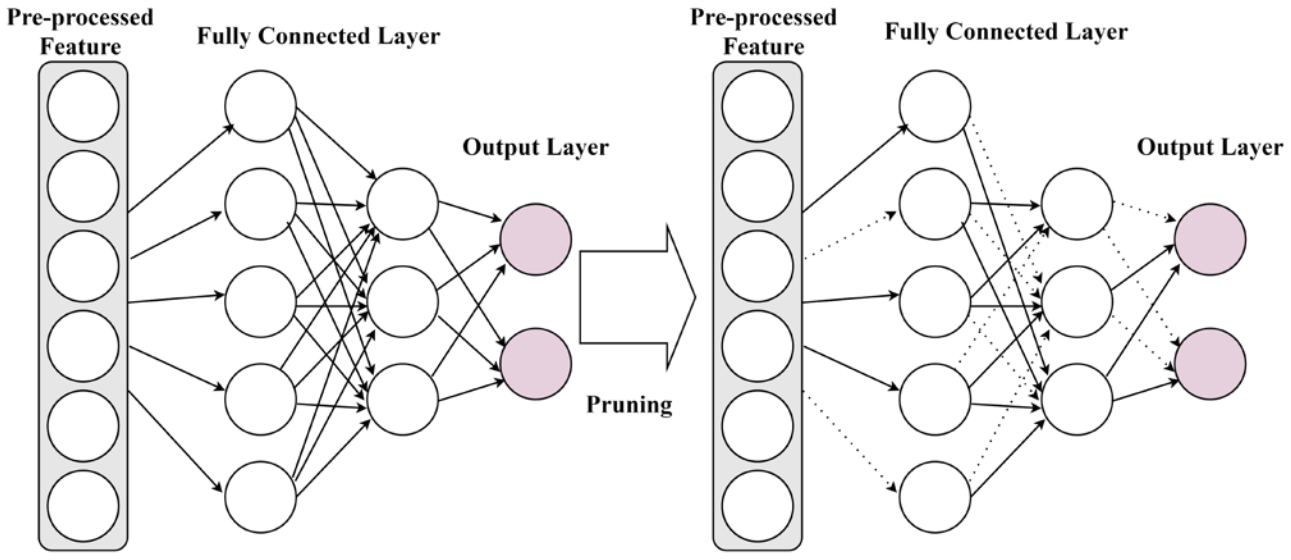

Fig. 3. Example of deep neural network pruning.

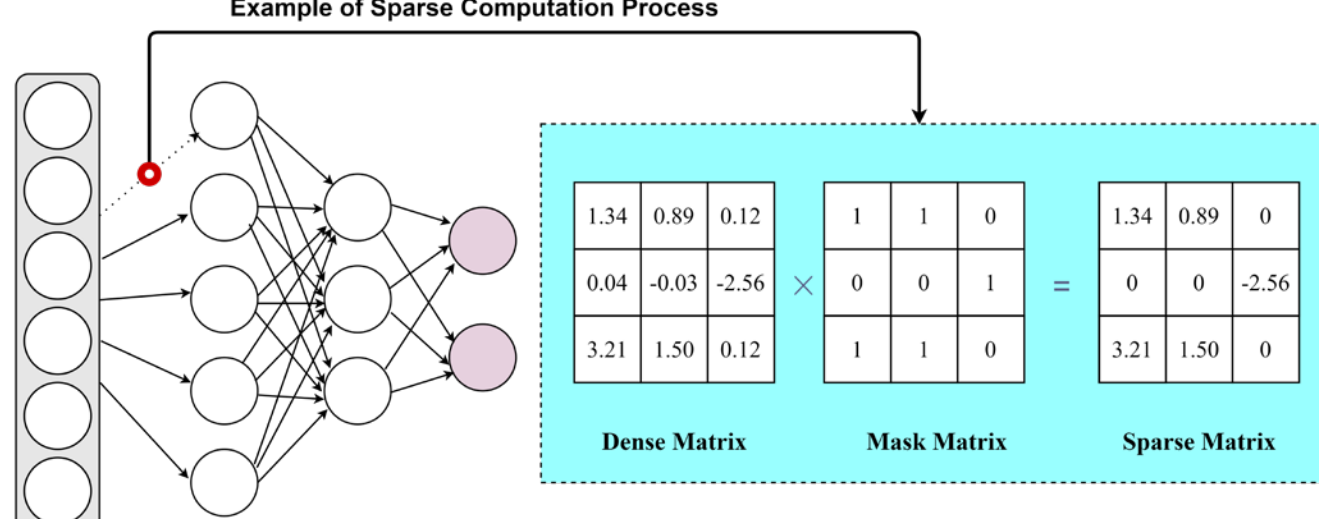

Fig. 4. Computation on sparse neural network.

In this work, we apply a masking technique to selectively prune network connections during the training phase [24]. Initially, the network is set up and pre-trained to focus on predicting frequency metrics. We then introduce a binary mask tailored to match the dimensions of the weight matrix in each pruned layer. This mask determines which weights are active during the forward execution phase, as shown in Fig. 4. The weights are organized by their absolute values, and those with the smallest values are deactivated, thereby reducing the network's computational demands and simplifying its structure. The parameter matrix's sparsity is gradually increased from an initial value of $s_0$ to a final value $s_{final}$, cross $\mu$ pruning steps, with a pruning frequency $\Delta e$:

$$s_e = s_{final} + \left(s_0 - s_{final}\right)\left(1 - \frac{e - e_0}{\mu \Delta e}\right)^3 \tag{26}$$

$$\text{for } e \in \{e_0, e_0 + \Delta e, \dots, e_0 + \mu \Delta e\}$$

This pruning scheme can allow the neural network model to keep critical connections while eliminating redundant weights.

To maintain the accuracy of the network despite increased sparsity, we progressively apply the same type of binary masks during back-propagation. This approach ensures that weights that are turned off during the forward pass do not change during the reverse pass, thereby preserving consistency in the training cycle and minimizing any potential loss of accuracy due to the pruning process.

A comparison between three sparsity techniques is listed in TABLE II. It could be observed that the cubic delay strategy mitigates abrupt performance degradation by prioritizing weight pruning in less critical network segments.

TABLE II.
Comparison of prediction accuracy [%] of different sparsification algorithms

| Sparsity | 0% | 20% | 40% | 60% | 80% |
|---|---|---|---|---|---|
| Exponential Decay | 93.15 | 92.56 | 92.76 | 86.78 | 25.66 |
| Linear Decay | 93.15 | 91.87 | 89.12 | 35.54 | 24.19 |
| Cubic Decay | 93.15 | 92.93 | 92.64 | 88.75 | 87.47 |

## B. Active ReLU Linearization

Directly adopting the ReLU activation function from neural networks into the framework of MILP such as SCUC will lead to nonlinearity. Such nonlinearity could be eliminated with piecewise linearization that would still result in a significant increase in binary variables [25], substantially amplifying the computational burden and reducing the model efficiency. Linear approximation of ReLU activation function has been considered as an effective method to handle such issues [26]. However, such approximation often involves simplifying the activation function, which can result in loss of critical nonlinear behavior that trained DNNs capture.

To address this issue, an innovative method that actively linearizes the ReLU function has been introduced. This approach is designed to streamline the complexity of DNN models while maintaining a high level of prediction accuracy with minimal reduction.

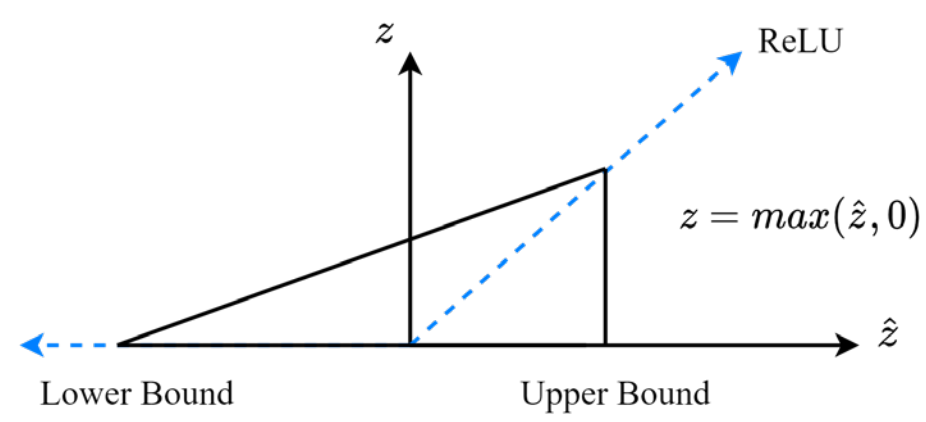

Fig. 5. The linear approximation of ReLU activation function.

Fig. 5 illustrates the approximation approach for the ReLU function, where $\hat{z}$ represents the weighted sum of input signals to the node, and $z$ denotes the activated output of the node. By utilizing the upper bound (UB) and lower bound (LB) of $\hat{z}$, the relationship between $z$ and $\hat{z}$ can be approximated through a set of constraints: $z \geq 0$ , $z \geq \hat{z}$ , and $z \leq \mathrm{UB} \cdot (\hat{z} - \mathrm{LB})/(\mathrm{UB} - \mathrm{LB})$. The UB and LB can be obtained by sending all the samples in the training dataset to the trained ML model and record the maximum and minimum pre-activated ReLU values for each neuron.

However, the implementation of such ReLU linearization for all neurons may introduce large approximation errors into the SCUC model, potentially causing the frequency security constraints to become invalid. To address this concern, we propose an active selection method aimed at minimizing the approximation error while maintaining the valid nature of frequency dynamic security constraints. A nodal positivity index $\varepsilon_{q[l]}$ is developed to estimate the percentage of positive pre-activated values of neuron node $l$ in the $q$-th layer.

$$\varepsilon_{q[l]} = \frac{1}{N_S}\left(\sum_{s \in NS} \hat{z}_{q[l],s} - \sum_{s \in NS}\left|\hat{z}_{q[l],s} - \frac{1}{N_S}\sum_{s \in NS} \hat{z}_{q[l],s}\right|\right) \geq \gamma \tag{27}$$

where $\gamma$ is the requirement threshold being set to select neuron nodes that suitable for ReLU linearization. For the prediction of a given case at hour $t$, (24a)-(24e) for ReLU function in selected neurons can be replaced by (28a)-(28c) as follows,

$$z_{q[l],t} \geq \hat{z}_{q[l],t}, \forall q, \forall l, \forall t, \tag{28a}$$

$$z_{q[l],t} \leq \frac{\mathrm{UB}_{q[l]} \cdot (\hat{z}_{q[l],t} - \mathrm{LB}_{q[l]})}{\mathrm{UB}_{q[l]} - \mathrm{LB}_{q[l]}}, \forall q, \forall l, \forall t, \tag{28b}$$

$$z_{q[l],s} \geq 0, \forall q, \forall l, \forall s, \tag{28c}$$

### C. Active Data Sampling Approach

Adjusting the predictor to focus on data close to critical thresholds through specific active sampling can lead to inaccuracies in the network's initial outputs. As a consequence, the frequency security constraints that depend on this data may not remain fully effective. A crucial consideration is maintaining a balance between reducing the number of parameters (sparsity) and minimizing approximation errors. We advocate for implementing an active data sampling strategy to improve the SCUC model's robustness and efficacy.

The task of assigning RoCoF values to a large collection of samples to meet specific distribution goals is challenging, often termed the labeling bottleneck [15]. To tackle this issue and generate an adequate amount of accurately labeled training data, we utilize an active sampling method. This process starts with the creation of a dataset via model-based methods, known as dataset $\mathcal{F}$. Each sample in dataset $\mathcal{F}$ is then precisely labeled with the highest frequency deviations and nodal RoCoF values, averaged over a 0.1-second interval following the most critical contingency event, to ensure the dataset captures essential dynamic responses for effective model training. The labeled dataset $\mathcal{F}$ is then assigned a security label, denoted as $f_{sec}$. If both $\Delta f_{max}$ and $\dot{f}_{max}$ of a sample fall within the security range, $f_{sec}$ is labeled as 1. Conversely, if any of the frequency metrics violates the threshold, $f_{sec}$ is labeled as 0. Finally, a frequency security discriminator is trained using the labeled dataset $\mathcal{F}$.

The active sampling approach proposed here is utilized on the unlabeled dataset $\mathcal{U}$ , leveraging the well-trained discriminator. Subsequently, samples lacking posterior probabilities from the discriminator are selected for training the multi-state predictor related to frequency. The sampling strategy relies on entropy as a metric proposed in [27]:

$$x_{sec} = \underset{\mathcal{U}}{\operatorname{argmax}} \left( -\sum_{1}^{\mathcal{U}} p(y_i|\mathcal{U}) \log p(y_i|\mathcal{U}) \right) \tag{29}$$

$$x_{insec} = \underset{\mathcal{U}}{\operatorname{argmin}} \left( -\sum_{1}^{\mathcal{U}} p(y_i|\mathcal{U}) \log p(y_i|\mathcal{U}) \right) \tag{30}$$

where $y_i$ spans all potential labels, with $p(y_i|\mathcal{U})$ representing the predicted posterior probability of belonging to class $y_i$. This sorting procedure identifies the maximum and minimum values from the finite set of numerical values. The selected $x_{sec}$ represents samples that closely distribute around the threshold, thereby enhancing the prediction accuracy. While $x_{insec}$ denotes samples aiding the output of the fine-tuned frequency metrics predictor to closely approximate the original.

## V. Mixed-Integer Formulations of ALSNN-SCUC

In this section, we formulate the proposed ALSNN-SCUC model, considering DNN-based frequency security constraints. The objective of the modified ALSNN-SCUC model is to minimize the overall operating cost while subjecting to various system operational constraints and ensuring the stability of system frequency. The formulation is presented below:

$$\min_{\Phi} \sum_{g \in G} \sum_{t \in T} (c_g P_{g,t} + c_g^{NL} u_{g,t} + c_g^{SU} v_{g,t} + c_g^{RE} r_{g,t}) \tag{31a}$$

$$\sum_{g \in G(n)} P_{g,t} + \sum_{k \in K(n-)} P_{k,t} - \sum_{k \in K(n+)} P_{k,t} - D_{n,t} + E_{n,t} = 0, \quad \forall n,t \tag{31b}$$

$$P_{k,t} - b_k(\theta_{n,t} - \theta_{m,t}) = 0, \ \forall k,t \tag{31c}$$

$$-P_k^{max} \le P_{k,t} \le P_k^{max}, \ \forall k,t \tag{31d}$$

$$P_g^{min} u_{g,t} \le P_{g,t}, \quad \forall g,t \tag{31e}$$

$$P_{g,t} + r_{g,t} \le u_{g,t} P_g^{max}, \quad \forall g,t \tag{31f}$$

$$0 \le r_{g,t} \le R_g^{re} u_{g,t}, \quad \forall g,t \tag{31g}$$

$$\sum_{j \in G} r_{j,t} \ge P_{g,t} + r_{g,t}, \quad \forall g,t \tag{31h}$$

$$P_{g,t} - P_{g,t-1} \le R_g^{hr}, \quad \forall g,t \tag{31i}$$

$$P_{g,t-1} - P_{g,t} \le R_g^{hr}, \quad \forall g,t \tag{31j}$$

$$v_{g,t} \ge u_{g,t} - u_{g,t-1}, \ \ \forall g,t \tag{31k}$$

$$v_{g,t+1} \le 1 - u_{g,t}, \ \forall g, t \le nT - 1 \tag{31l}$$

$$v_{g,t} \le u_{g,t}, \quad \forall g,t \tag{31m}$$

$$\sum_{s=t-UT_g}^{t} v_{g,s} \le u_{g,t}, \forall g, t \ge UT_g \tag{31n}$$

$$\sum_{s=t-UT_g}^{t+DT_g} v_{g,s} \le 1 - u_{g,t}, \forall g, t \ge nT - DT_g \tag{31o}$$

$$u_{g,t}, v_{g,t} \in \{0,1\}, \quad \forall g,t \tag{31p}$$

The objective function of ALSNN-SCUC model is detailed in equation (31a), with foundational constraints spanning equations (31b) to (31o). Equation (31b) ensures the balance of power at each node, while equation (31c) dictates the flow of power across the network, subject to the transmission limits specified in equation (31d). Constraints on the scheduled production and reserve capacities of generators, which are limited by their maximum generation capacity and ramping capabilities, are set out in equations (31e) to (31j). Reserve requirements, which guarantee adequate reserves to manage the loss of a single generator, are established in equation (31h). The operational statuses of conventional units, including start-up and shutdown rules, are managed by binary variables, with specific guidelines for minimum downtime and uptime provided in equations (31n) and (31o) respectively. Additionally, binary variables that indicate generator start-up and commitment status are outlined in equation (31p).

Due to the inclusion of the max operator in the disturbance feature vector $\varpi_t^G$, direct utilization in the encoding formulations is impractical. Thus, we introduce an auxiliary variable to signify whether the dispatched generation of generator $g$ is the largest in the scheduling period t. The reformulations are detailed as follows,

$$P_{g,t}^{con} - P_{g,t} \le A(1 - \mu_{g,t}), \quad \forall g,t \tag{32a}$$

$$\sum_{g \in G} \mu_{g,t} = 1, \quad \forall t \tag{32b}$$

$$\mu_{g,t} \in \{0,1\}, \quad \forall g,t \tag{32c}$$

Let $A$ represent a large constant number. Equation (32a) mandates $\mu_{g,t}$ is set to zero if the dispatched power of any generator $P_{g,t}^{con}$ exceeds that of the interested generator $g$ at period t. Equation (32b) guarantees that only one generator is designated as the potential largest contingency. The combination of equations (32a) and (32b) ensures that generator $g$ possesses the highest output power, resulting in $\mu_{g,t}$ being assigned the value of 1. To further incorporate both the magnitude and spatial aspects of the disturbance into the feature vector, variable $\rho_{g,t}$ is introduced, where $\rho_{g,t}$ equals to the largest generation at period t, the corresponding constraints can be formulated as follows,

$$\rho_{g,t} - P_{g,t} \geq -A(1-\mu_{g,t}), \qquad \forall g,t \tag{33a}$$

$$\rho_{g,t} - P_{g,t} \leq A(1-\mu_{g,t}), \qquad \forall g,t \tag{33b}$$

$$0 \leq \rho_{g,t} \leq A\mu_{g,t}, \qquad \forall g,t \tag{33c}$$

Thus, the overall feature vector of a case $x_t$ can be then defined as follows,

$$x_t = [u_{1,t},\cdots,u_{N_G,t},\ \rho_{g,t},\cdots,\ \rho_{N_G,t}, P_{1,s},\cdots,P_{N_G,s}] \tag{34}$$

Then we outline the formulation of the proposed DNN-based predictor for frequency related events. As previously mentioned, neurons in each layer with a positivity index $\varepsilon_{q[l]}$ greater than $\gamma$ are isolated and included into set $\mathcal{H}$. The adjusted ReLU activation for these selected neurons is articulated as follows.

$$z_1 = x_t W_1 + b_1, \forall t, \tag{35a}$$

$$z_{q[l],t} \geq \hat{z}_{q[l],t}, \forall q, \forall l \in \mathcal{H}, \forall t, \tag{35b}$$

$$z_{q[l],t} \leq \frac{UB_{q[l]} \cdot (\hat{z}_{q[l],t} - LB_{q[l]})}{UB_{q[l]} - LB_{q[l]}}, \forall q, \forall l \in \mathcal{H}, \forall t, \tag{35c}$$

$$z_{q[l],t} \leq \hat{z}_{q[l],t} - A(1-a_{q[l],t}), \forall q, l \in \bar{\mathcal{H}}, t, \tag{35d}$$

$$z_{q[l],t} \geq \hat{z}_{q[l],t}, \forall q, \forall l \in \bar{\mathcal{H}}, \forall t, \tag{35e}$$

$$z_{q[l],t} \leq A a_{q[l],t}, \forall q, \forall l \in \bar{\mathcal{H}}, \forall t, \tag{35f}$$

$$z_{q[l],t} \geq 0, \forall q, \forall l, \forall t, \tag{35g}$$

$$a_{q[l],t} \in \{0,1\}, \forall q, \forall l, \forall t, \tag{35h}$$

$$z_{N_L,t} W_{N_L+1}^{dev} + b_{N_L+1}^{dev} \leq f_{\text{norm}} - f_{\text{lim}} \tag{35i}$$

$$z_{N_L,t} W_{N_L+1}^{rcf} + b_{N_L+1}^{rcf} \leq -\text{RoCoF}_{\text{lim}} \tag{35j}$$

## VI. Result Analysis

To validate the effectiveness of the ALSNN-SCUC, case studies are conducted on IEEE 24-bus system [28]. Data generation relied on mathematical models implemented in Python using Pyomo [29]. While the time-domain simulation and labeling processes were carried out using PSS/E [30], SCUC models are performed using Pyomo on a Windows laptop with Intel(R) Core(TM) i7 2.60GHz CPU and 16 GB RAM.

### *A. Predictor Training*

The data-driven approach necessitates a dataset, denoted as $\mathcal{L}$, comprising 3000 labeled samples for training the frequency security discriminator. Monte Carlo simulation is subsequently employed to generate 10,000 samples of load profiles and RES forecasts from a Gaussian distribution. It's worth noting that if historical data on uncertainty is available, it can be utilized as the training dataset to inform the optimization models.

*1) Distribution of different dataset:* The AS method is employed to select appropriate samples from the unlabeled dataset. Additionally, two benchmark datasets are established: (1) a random dataset generated using the randomly selected (RS) method, and (2) a dataset based on the uncertainty sampling (US) strategy [15] , which selects instances with a posterior probability of being positive closest to 0.5.

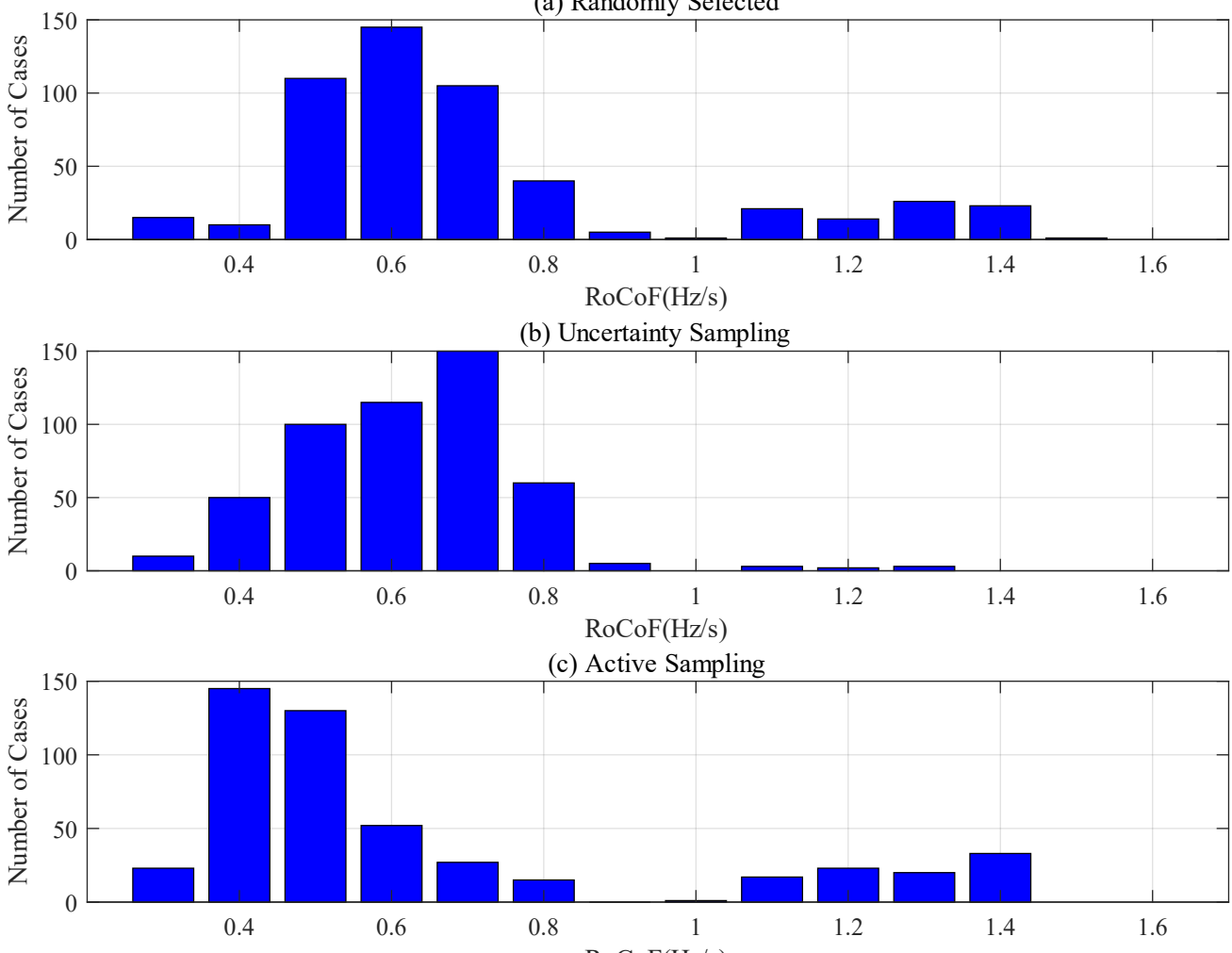

Fig. 6. RoCoF value distributions for three different datasets.

Figure 6 illustrates the distribution of peak RoCoF values in the sampled scenarios. Using the RS method, most cases cluster between 0.5 and 0.8 Hz/s, with fewer than 2% approaching the lower range of 0.4 to 0.5 Hz/s near the threshold. In contrast, the US method shows about 40% of cases aligning closely with the threshold. It is important to note that samples exhibiting RoCoF values above 1.0 Hz/s are omitted, which may lead to non-binding constraints within the SCUC model.

2) *Validation of different predictor:* Tables III and IV present the outcomes of the DNN-based predictor trained on various datasets. Across all cases, the RoCoF prediction accuracies surpass 93.27%, with an error tolerance of 5%. Particularly, the predictor trained with the active sampling dataset achieves the highest prediction accuracy for both RoCoF and frequency deviation nadir, as expected, providing more accurate constraints.

TABLE III. ROCOF validation accuracy [%]

| Tolerance | 10% | 9% | 8% | 7% | 6% | 5% |
|---|---|---|---|---|---|---|
| RS Dataset | 98.39 | 97.92 | 97.27 | 96.05 | 94.66 | 93.27 |
| US Dataset | 99.26 | 98.26 | 97.93 | 97.23 | 96.57 | 96.29 |
| AS Dataset | 98.44 | 98.07 | 97.60 | 97.27 | 96.28 | 95.11 |

An extensive out-of-sample dataset is employed to test the efficacy of the sparse neural network. Sparsity levels are methodically varied from 0% to 90%, increasing by 10% at each step. Figure 7 shows the prediction accuracy within a 5% margin. At 0% level, the AS dataset achieves the highest RoCoF prediction accuracy of 93.15%, outperforming random sampling (91.18%) and uncertainty sampling (92.53%). Interestingly, the prediction accuracy for RoCoF using the US dataset declines sharply as sparsity reaches 60%, suggesting reduced model robustness. In contrast, the predictor using the RS dataset shows a decrease in accuracy to 80.04% at 50% sparsity and experiences a complete decline in performance at 70% sparsity. However, the predictor utilizing the AS dataset markedly excels, maintaining superior performance of 87.47% prediction accuracy even at an 80% sparsity level.

TABLE IV. Frequency deviation nadir validation accuracy [%]

| Tolerance | 10% | 9% | 8% | 7% | 6% | 5% |
|---|---|---|---|---|---|---|
| RS Dataset | 97.96 | 97.44 | 96.57 | 93.71 | 89.50 | 84.76 |
| US Dataset | 99.01 | 98.73 | 97.79 | 96.62 | 93.36 | 91.36 |
| AS Dataset | 98.88 | 98.63 | 97.17 | 95.18 | 91.16 | 88.78 |

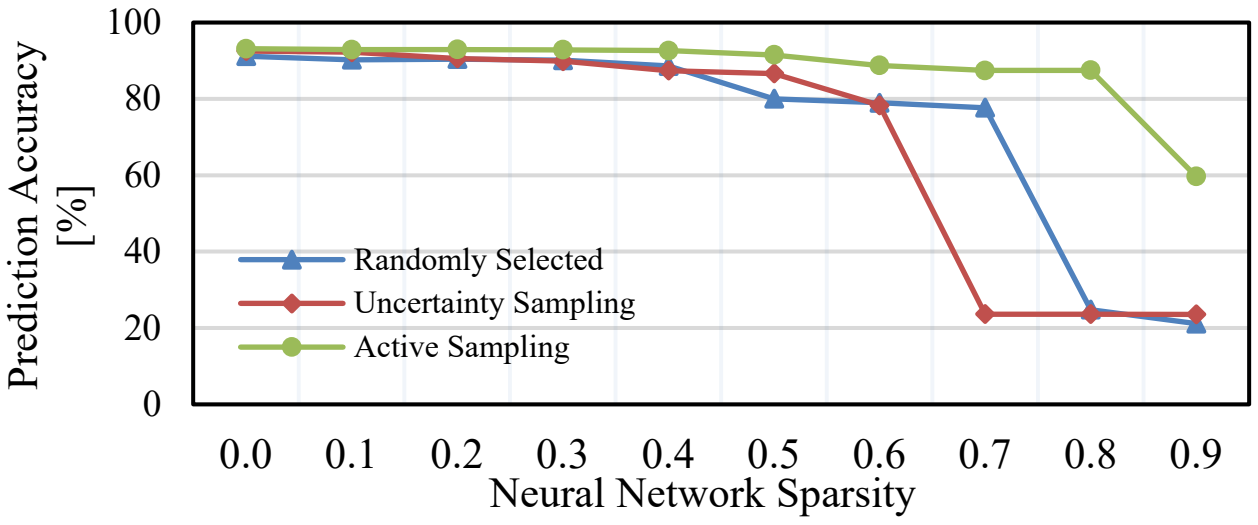

Fig.7. RoCoF prediction accuracy with different DNN sparsity.

The findings illustrated in Fig. 8 reveal that the predictor trained on the active sampling dataset demonstrates superior robustness against sparsity in comparison to the other two cases. Unlike the RoCoF prediction accuracy, the frequency deviation prediction accuracy proves to be more susceptible to alterations in network sparsity. Despite its superiority over the other predictors, the accuracy of the predictor trained on the active sampling dataset diminishes at a sparsity of 0.7.

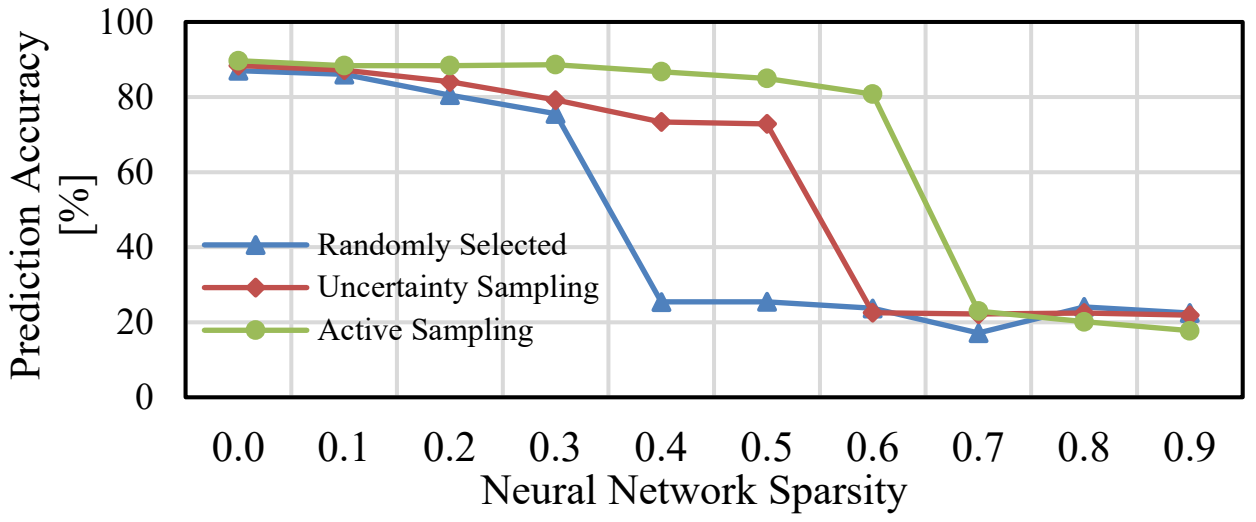

Fig.8. Frequency deviation prediction accuracy with different DNN sparsity.

The proposed ML-based frequency-metrics predictor can achieve the above-reported accuracy through a synergistic combination of model architecture, feature engineering, and data processing strategies. For instance, the proposed active sampling strategy prioritizes threshold-near samples, which greatly improves the predictor's performance when it is incorporated in SCUC with increased sensitivity, around the threshold, to frequency-related constraint binding.

It should also be noted that the ML model prediction accuracy may drop sharply beyond a certain high sparsity level such as 90% for RoCoF prediction when using the active sampling method in this paper. This is due to the significant reduction in weight redundancy and in ML model representational capacity at very high sparsity levels. To avoid using overly high sparsity levels that risk catastrophic performance loss, a recommended safeguard, as adopted in this paper, would be to implement staged sparsification with small sparsity incremental steps, gradually increasing the sparsity level until the desired prediction accuracy cannot be met. Then, as illustrated in Fig. 1, we can integrate physical grid system simulation accuracy and stability monitoring into the pipeline, ensuring proper interactions between the ML model and grid system physical performance; we can thus further adjust or lower the sparsity level if prediction errors or stability indicators exceed predefined thresholds. Finally, case-specific sparsity limits based on grid size and RES penetration should be documented to guide practitioners away from using overly sparse ML models that risk over-pruning. These measures leverage our findings that accuracy collapse can be avoided through careful design, ensuring that sparsity is balanced with performance.

## B. *Simulation Results Validations*

In this section, we validate the proposed ALSNN-SCUC method. We establish the validation criteria based on post-contingency frequency constraints, where the maximal frequency deviation should not exceed 0.5 Hz, and the RoCoF must maintain a value higher than -0.5 Hz/s to prevent the activation of RoCoF-sensitive protection relays. To ensure rigorous validation, we set the MIP gap to 0.1% for the solver.

*1) Results of Different Sparsity Levels:* We first investigate the impact of sparsity on the efficiency of Sparse Neural Network-based security-constrained unit commitment (SNN-SCUC) models without active ReLU approximation, employing a scheduling horizon of 4 hours. The heatmap displayed in Fig. 9 showcases the SCUC solving times. Notably, at 80% sparsity, the computational times for both RS and US scenarios drop to less than 5 seconds, indicating that the frequency-related constraints are not binding. However, these scenarios do not guarantee frequency stability, as evidenced by time-domain simulations. In contrast, the AS-based predictor exhibits significantly greater robustness when integrated into the MILP framework, with constraints remaining non-binding even at 90% sparsity.

As shown in Table V, the NN-based frequency constraints in SNN-SCUC and ALSNN-SCUC models with different sparsifications are all binding, indicating the effectiveness of these constraints in ensuring stable frequency performance. Both Table V and Fig. 9 show that the adopted DNN sparsification strategy can substantially reduce the SCUC computing time, as well as the proposed active ReLU linearization strategy at various SNN sparsity levels.

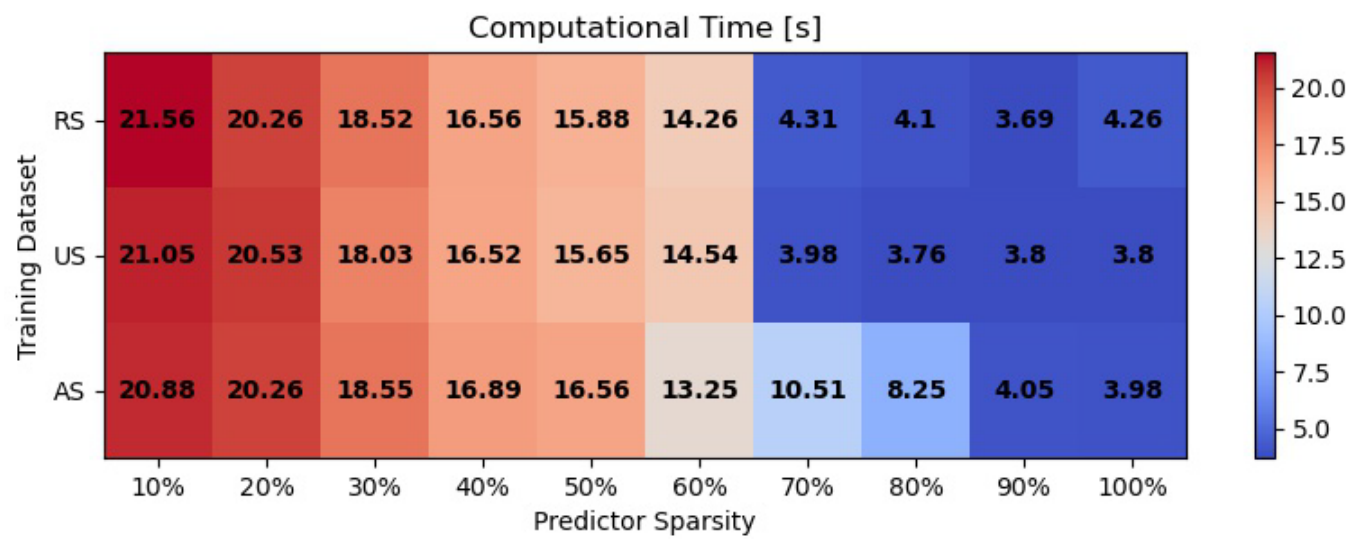

Fig. 9. Computational time of SNN-SCUC model with different sparsity level.

TABLE V. Computational Time of Various SCUC Models

| Sparsity | 0% | 20% | 40% | 60% | 80% |
|---|---|---|---|---|---|
| SNN-SCUC | 21.23 | 20.26 | 16.89 | 13.25 | 8.25 |
| ALSNN-SCUC | 17.15 | 15.33 | 12.79 | 10.22 | 7.95 |
| Bindingness of NN-based constraints | Yes | Yes | Yes | Yes | Yes |

Tables VI and VII present the detailed effects of sparsification and active linearization on MILP complexity respectively. Quantitative experiments demonstrate that both sparsification and active linearization improve computational efficiency while preserving the core problem structure. Sparsification from 0% to 80% significantly reduces non-zero elements (3%, 66,355 to 64,403) and computational time (63.4%, 21.23s to 8.25s), while the numbers of constraints, total variables, and binary/continuous variables remain stable, proving efficiency gains are independent of problem definition changes. Similarly, active linearization maintains stable constraint counts and variable structures but reduces solve time by transforming nonlinear components, confirming it enhances efficiency without altering the fundamental problem framework.

TABLE VI.
Summary of SNN-SCUC Models on the IEEE 24-bus System

| Sparsity | 0% | 20% | 40% | 60% | 80% |
|---|---|---|---|---|---|
| Constraints | 18,602 | 18,602 | 18,602 | 18,602 | 18,602 |
| Variables | 5,704 | 5,704 | 5,704 | 5,704 | 5,704 |
| Binary Variables | 2,856 | 2,856 | 2,856 | 2,856 | 2,856 |
| Continuous Variables | 2,848 | 2,848 | 2,848 | 2,848 | 2,848 |
| Non-zeros | 66,355 | 65,881 | 65,435 | 64,927 | 64,403 |
| Computational Time [s] | 21.23 | 20.26 | 16.89 | 13.25 | 8.25 |

TABLE VII.
Summary of ALSNN-SCUC Models on the IEEE 24-bus System

| Sparsity | 0% | 20% | 40% | 60% | 80% |
|---|---|---|---|---|---|
| Constraints | 12,755 | 12,738 | 12,738 | 12,746 | 12,746 |
| Variables | 5,144 | 5,144 | 5,148 | 5,148 | 5,148 |
| Binary Variables | 1,772 | 1,772 | 1,772 | 1,775 | 1,775 |
| Continuous Variables | 3372 | 3372 | 3372 | 3372 | 3372 |
| Non-zeros | 62,122 | 61,551 | 61,027 | 59,523 | 58,965 |
| Computing Time [s] | 17.15 | 15.33 | 12.79 | 10.22 | 7.95 |

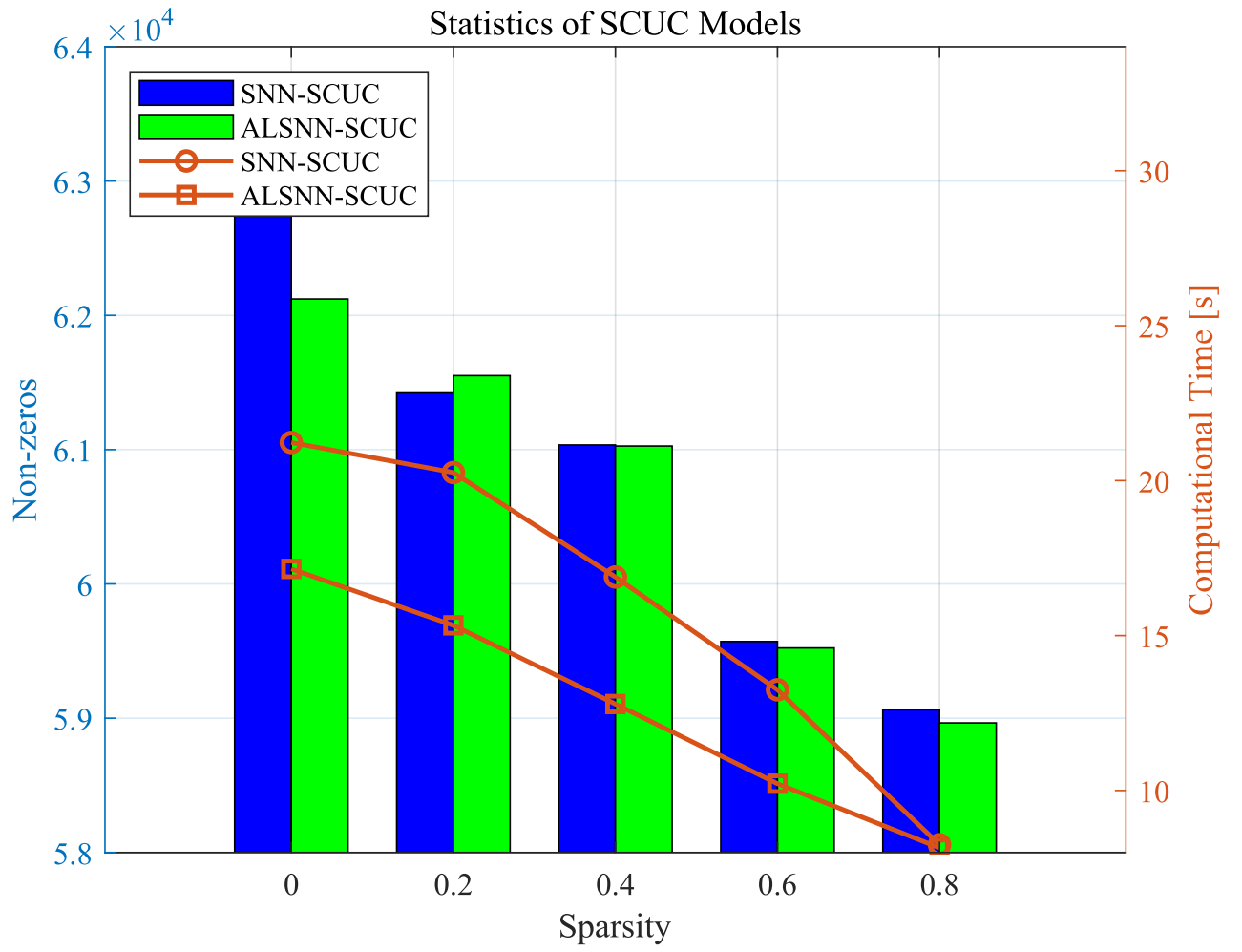

Fig. 10. Statistics of AS dataset-based SCUC models.

Figure 10 represents the number of non-zero elements and the computational time of SNN-SCUC and ALSNN-SCUC models. It is notable that there exists a close correlation between the number of non-zero elements and the computational time of the SNN-SCUC model. As sparsity level increases, the number of non-zero elements within both SNN-SCUC and ALSNN-SCUC models decreases, leading to reduced computational time. This indicates that active linearization has a positive impact on the computational efficiency of the MILP problem while keeping the problem's constraint and variable structures stable.

*2) Validation of ALSNN-SCUC:* In this section, we evaluate the performance of the proposed ALSNN-SCUC model. The sparsity of the predictor is determined based on the approach illustrated in Figure 1. The total scheduling horizon for the test case spans 24 hours, with hours 9 to 12 designated as the time instance where frequency-related constraints are enforced. Leveraging the sparse computation algorithm alongside the active ReLU approximation method, the sparse predictor achieves a nodal RoCoF prediction accuracy of 87.47% and a frequency deviation prediction accuracy of 80.80%. The threshold $\gamma$ for active neuron selection is set to 0.25. $\dot{f}_{max}$ and $\Delta f_{max}$ are through time domain simulations using PSS/E at hour 10, considering G-1 as worst contingency scenario.

The total UC costs and time domain simulation outcomes for all cases are summarized in Table VIII. The integration of frequency-related constraints leads to additional costs incurred in ensuring the stability of the frequency nadir and nodal RoCoF. ERC-SCUC has the least cost compared to all other frequency-constrained SCUC models; however, it causes severe frequency stability violations since it enforces a system-wide RoCoF and fails to capture nodal frequency responses. The solution derived from the LRC-SCUC model tends to be overly conservative, primarily due to approximation errors inherent in the derivation of the second-order model. Notably, a considerable increase in computational time is observed in cases where no ReLU approximation is employed in the DNN-SCUC. In contrast, the proposed ALSNN-SCUC model achieves a notable reduction in computational time, decreasing to 29.56s from DNN-SCUC's 4292.18s, a reduction of over 99%. Compared to the SNN-SCUC model, the proposed ALSNN-SCUC model can still reduce the computing time by 67.6%, which demonstrates the effectiveness of the proposed active ReLU linearization strategy in accelerating the SCUC solving process. All the three learning-based models, DNN-SCUC, SNN-SCUC, and ALSNN-SCUC, which enforces nodal frequency performance, are proved to have high model reliability in terms of meeting the pre-specified frequency requirements. It should be also noted that $\Delta f_{max}$ of RoCoF constrained models are all within the safe range, implying RoCoF related constraints are more likely to bind comparing to the frequency deviation constraints in low inertia system. Inertia related protections would be a main factor that limits the transition toward RES dominant system.

TABLE VIII. Comparison of different SCUC models on IEEE 24-bus system

| Model | Total Cost [$] | Computational Time [s] | $\dot{f}_{max}$ [Hz/s] | $\Delta f_{max}$ [Hz] |
|---|---|---|---|---|
| T-SCUC | 523,569 | 6.89 | -1.10 | 0.51 |
| ERC-SCUC | 528,822 | 7.33 | -0.69 | 0.29 |
| LRC-SCUC | 547,776 | 18.35 | -0.44 | 0.22 |
| DNN-SCUC | 531,233 | 4,292.18 | -0.50 | 0.24 |
| SNN-SCUC | 531,457 | 91.17 | -0.50 | 0.24 |
| ALSNN-SCUC | 531,430 | 29.56 | -0.50 | 0.24 |

Figures 11 and 12 illustrate the system uniform RoCoF responses and nodal RoCoF responses for different scheduling cases, respectively. It is observed that the system uniform RoCoF remains within the safe range for both scenarios, with the ERC-SCUC model providing the optimal solution in terms of uniform RoCoF security. A combined analysis of Figures 11 and 12 reveals that although the system uniform RoCoF does not exceed the threshold for ERC-SCUC and LRC-SCUC models, the stability of nodal RoCoF at several generator buses cannot be guaranteed. In contrast to the LRC-SCUC model, which employs model-based constraints, data-driven methods effectively ensure nodal RoCoF stability. The highest RoCoF recorded in the proposed ALSNN-SCUC model is -0.50 Hz/s, demonstrating that the SCUC solution satisfies the threshold without any conservatism.

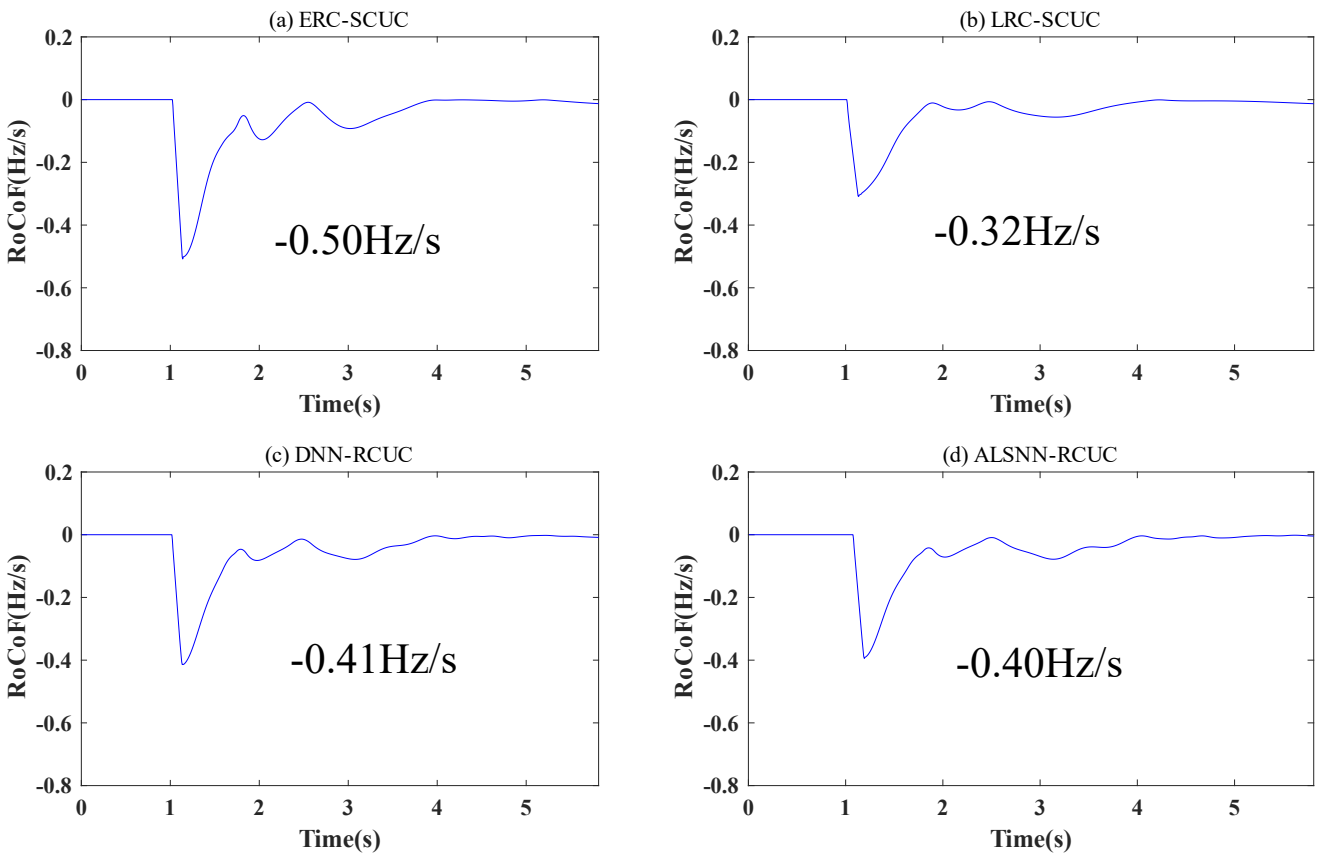

Fig. 11. Uniform RoCoF evolutions under worst contingency at hour 10.

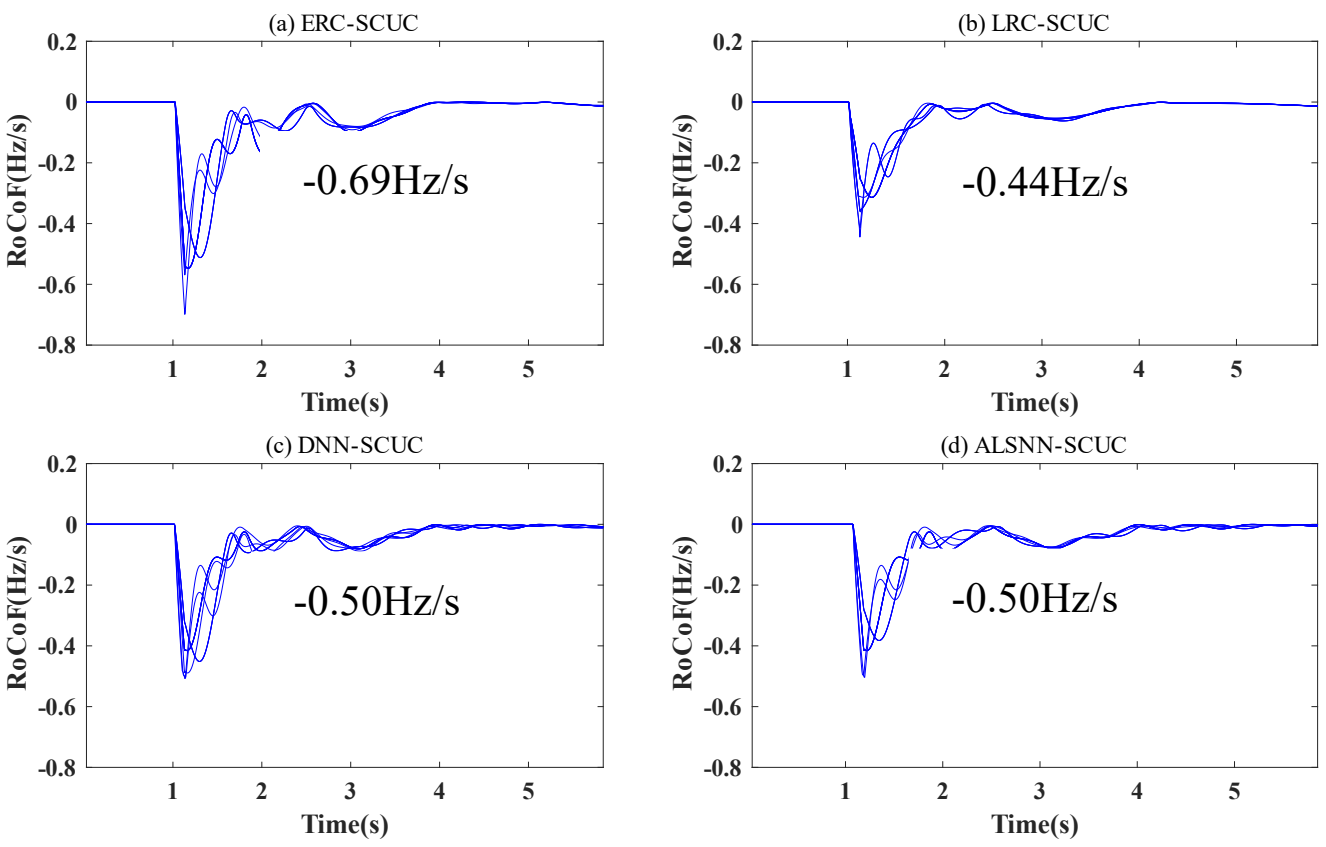

Fig. 12. Nodal RoCoF evolutions under worst contingency at hour 10.

## *C. Scalability Studies on the IEEE 118-Bus System*

To validate scalability, the proposed ALSNN-SCUC model has been tested on the larger IEEE 118-bus system. Periods 18-21 are constrained with DNN-based frequency security constraints. Table IX quantifies the computational efficiency gains achieved through sparsification. As the sparsity level increases from 0% to 80%, the number of non-zero elements decreases by 10.01% from 261,288 to 235,131, directly reducing the model complexity. This sparsity-driven compression significantly accelerates the solving process and save the computing time by 95.3% while preserving the original problem structure.

Table X compares the dynamic and economic performance across various SCUC models. The T-SCUC solution incurred severe frequency violations, confirming traditional models' inadequacy in low-inertia systems. While ERC-SCUC and LRC-SCUC improved dynamic stability, the ERC-SCUC solution still leads to a RoCoF value beyond the pre-specified tolerance, and the LRC-SCUC solution, though meeting the frequency requirement, leads to an increased costs by 5.1%, due to its conservative approximation. On the other hand, the proposed ALSNN-SCUC model achieved optimal frequency security; it corresponds to a total cost of $1,618,783, which is even lower than DNN-SCUC. Critically, the proposed ALSNN-SCUC model can be solved much faster than DNN-SCUC, demonstrating that the proposed computational enhancement strategies, sparse computation and active linearization, can substantially relieve the computational burden while maintaining high-fidelity frequency security requirements.

TABLE IX. Results of ALSNN-SCUC models on the IEEE 118-bus system

| Sparsity | 0% | 20% | 40% | 60% | 80% |
|---|---|---|---|---|---|
| Constraints | 302,160 | 302,160 | 302,160 | 302,160 | 302,160 |
| Variables | 20,544 | 20,544 | 20,544 | 20,544 | 20,544 |
| Binary Variables | 4,624 | 4,624 | 4,624 | 4,624 | 4,624 |
| Continuous Variables | 12,520 | 12,520 | 12,520 | 12,520 | 12,520 |
| Non-zeros | 261,288 | 207,195 | 202,035 | 195,271 | 235,131 |
| Computing Time [s] | 5,461.94 | 2,834.43 | 1,026.50 | 415.97 | 256.46 |

TABLE X. Comparison of different SCUC models on IEEE 118-bus system

| Model | Total Cost [$] | Computational Time [s] | $\dot{f}_{max}$ [Hz/s] | $\Delta f_{max}$ [Hz] |
|---|---|---|---|---|
| T-SCUC | 1,553,296 | 45.65 | -1.42 | -0.35 |
| ERC-SCUC | 1,570,618 | 78.23 | -0.81 | -0.18 |
| LRC-SCUC | 1,633,142 | 661.34 | -0.40 | -0.12 |
| DNN-SCUC | 1,619,001 | 5,461.94 | -0.47 | -0.16 |
| SNN-SCUC | 1,621,137 | 557.19 | -0.45 | -0.16 |
| ALSNN-SCUC | 1,618,783 | 256.46 | -0.45 | -0.15 |

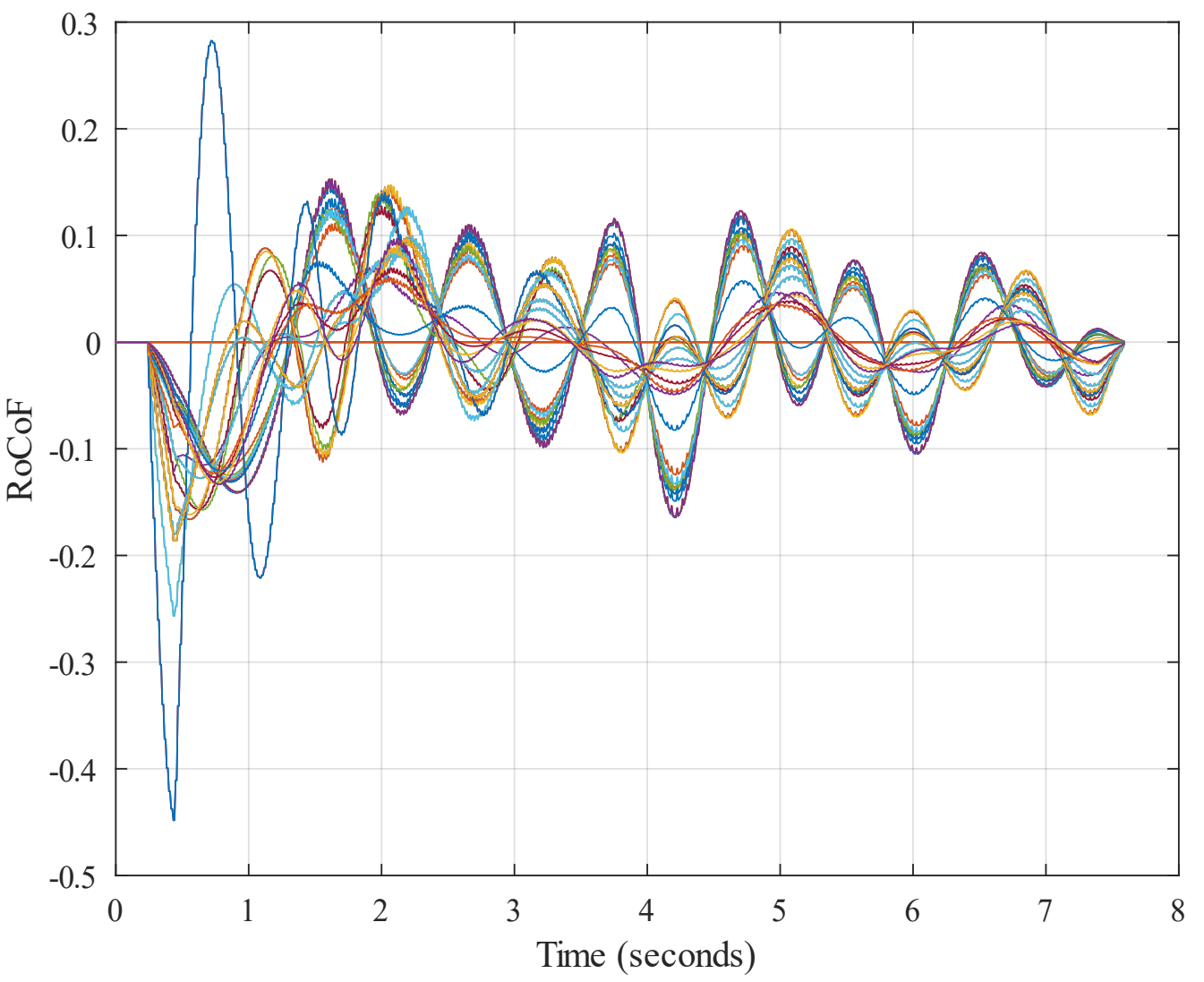

Fig. 13. Locational RoCoFs of the proposed ALSNN-SCUC model on the IEEE 118-bus system.

The locational RoCoF curves of the proposed ALSNN-SCUC model are plotted in Fig. 13. It can be observed that the proposed ALSNN-SCUC model can limit the maximum RoCoF to 0.45 Hz/s. In contrast, the RoCoF of physics-based ERC-SCUC model is 0.8 Hz/s greatly exceeding the threshold of 0.5 Hz/s, while LRC-SCUC provides overly conservative solutions. It should also be noted that oscillations in the IEEE 118-bus system are prominent, indicating that multi-area sys-tems may experience more complex post-contingency scenarios compared to smaller systems.

These results underscore the scalability of the proposed ALSNN-SCUC model. It enforces locational frequency stability constraints in large systems and can be solved to optimal solutions in a reasonable time, potentially enabling ALSNN-SCUC's practical deployment in real-world grids with high renewable penetration.

## VII. Conclusions and Future Work

### A. Conclusions

With growing RES in the power grid, maintaining system dynamic stability becomes increasingly critical. Since grid stability is largely affected by initial system conditions including generator status and output power, which are determined by look-ahead energy scheduling that traditionally focuses on steady-state power balancing and reliability only. To incorporate grid dynamic performance particularly frequency security constraints, this paper leveraged ML technologies to assist such modeling in SCUC and proposed an innovative ALSNN-SCUC model. A DNN-based ML model is developed to jointly predict both the maximum frequency deviation and maximum nodal RoCoF. To improve the robustness and accuracy of the ML-based predictor in conditions of sparse computation, an active sampling strategy is utilized. Furthermore, an active ReLU approximation is applied to strike a balance between computational efficiency and prediction accuracy. The results demonstrate the effectiveness of the proposed ALSNN-SCUC model in maintaining system frequency stability.

### B. Future Work

The ML-based frequency stability predictor can be extended to include other stability metrics with additional outputs of the neural network. The proposed method can also be extended to capture the impact of grid-forming inverters with virtual inertia. Future work will also explore extending such ML-assisted dynamics modeling and computational enhancement methods to real-time generation redispatch and grid control applications.

The grid topology and RES volatility are not explicitly encoded in this work; however, the predictor's robustness can be maintained through comprehensive offline training on dynamic cases that implicitly reflect topological and RES variations. While this work demonstrates the feature vector's adequacy for capturing essential dynamics, we will extend it and conduct additional tests by integrating grid topology reconfigurations and RES volatilities in the future. Regarding the data sample generation step, a potential future work can enhance it by considering the correlation between RES and load.

## Acknowledgment

This research was in part supported by the National Science Foundation (NSF) under Grant No. 2337598; and it was also partly supported by the NSF AI Institute for Advances in Optimization (Award 2112533). Any opinions, findings, and conclusions or recommendations expressed in this paper are those of the author(s) and do not necessarily reflect the views of the NSF.